\documentclass[12pt]{article}
\usepackage{amssymb}
\usepackage[pdftex,colorlinks=true,urlcolor=black,filecolor=black,linkcolor=black,
        pdftitle={A Fluid Generalization of Membranes.},
        pdfauthor={Mark D. Roberts},
        pdfsubject={QFT,strings,hydrodynamics},pdfkeywords={membrane,fluid,thermodynamics},
        pagebackref,pdfpagemode=None,bookmarksopen=true]{hyperref}
\newcommand{\be}{\begin{equation}}
\newcommand{\ee}{\end{equation}}
\newcommand{\bc}{\begin{center}}
\newcommand{\ec}{\end{center}}
\newcommand{\ber}{\begin{eqnarray}}
\newcommand{\ear}{\end{eqnarray}}
\newcommand{\ba}{\begin{array}}
\newcommand{\ea}{\end{array}}

\newcommand{\al}{\alpha}
\newcommand{\bt}{\beta}

\newcommand{\de}{\delta}

\newcommand{\dn}{\mu}

\newcommand{\el}{\ell}
\newcommand{\en}{\xi}

\newcommand{\ff}{\aleph}
\newcommand{\fr}{\frac}

\newcommand{\ga}{\gamma}
\newcommand{\hb}{\hbar}
\newcommand{\Hm}{{\cal H}}

\newcommand{\lb}{\label}

\newcommand{\Lg}{{\cal L}}
\newcommand{\n}{\nonumber\\}

\newcommand{\p}{\partial}
\newcommand{\ph}{\phi}
\newcommand{\pr}{\wp}

\newcommand{\rh}{\rho}
\newcommand{\Si}{\Sigma}
\newcommand{\si}{\sigma}
\newcommand{\sq}{\sqrt}
\newcommand{\ta}{\tau}

\newcommand{\Te}{\Theta}
\newcommand{\te}{\theta}

\begin{document}
\title{A Fluid Generalization of Membranes.}
\author{
\href{http://www.violinist.com/directory/bio.cfm?member=robemark}
{Mark D. Roberts},
54 Grantley Avenue,  Wonersh Park,  GU5 0QN,  UK,
}
\date{$24^{th}$ of March 2011}
\maketitle
\begin{abstract}
In a certain sense a perfect fluid is a generalization of a point particle.
This leads to the question as to what is the corresponding generalization for extended objects.
Here the lagrangian formulation of a perfect fluid is much generalized
by replacing the product of the co-moving vector which is a first fundamental form by
higher dimensional first fundamental forms;
this has as a particular example a fluid which is a classical generalization of a membrane;
however there is as yet no indication of any relationship between their quantum theories.
\end{abstract}
{\tiny\tableofcontents}
\section{Introduction.}\lb{intro}
There are three main types of dynamical system which have a lagrangian formulation:
fields theories,  extended objects,  and fluids.
Some other systems which have lagrangian formulation
include relative motion systems \cite{mdr42}.
Various relationships have been
found between string theories and field theories \cite{GSW}p.26-27,
also various string theories can be written as field theories \cite{tseytlin}.
Some relationships are known between extended objects and fluids
\cite{AS,j1,hoppe,GHY,GS},
these seem to be of the form of extended objects producing fluids.
In the present work it is fluids which are the more general object
which reduce to extended objects.
A generalization of string theory called M-theory is currently being sought,
whether there is any relationship between this hypothetical generalization and
fluids remains to be seen.
The prospect of generalizing the lagrangian theory of fluids rests on the observations that
a perfect fluid spacetime stress can be reduced to that of a congruence of point particles.
It seems that imperfect fluids do not often have a lagrangian description,
in particular heat conduction and anisotropic stress
do not seem to be derivable from a lagrangian \cite{mdr0910}.
The lagrangian theory of fluids involves all three of:
relativity, quantum theory and thermodynamics,
although there has not yet been any connection established with
quantum field theory on curved spacetime \cite{BD} which also involves all three.

Apart from the challenge of studying a formalism
which has contact with the above three theories,
there are at least {\bf four} other motivations for studying lagrangian fluid theory.
\begin{itemize}
\item
symmetry breaking,  see \cite{mdr25},
\item
low temperature physics,
so far there has been no contact at all of Lagrangian based theory with existing approaches,
such as \cite{CK},
\item
quantum cosmology,
with the geometry coupled to a fluid,  see for example \cite{bi:LR},
\item
the study of irreversible process,  see for example \cite{HL85},
a lagrangian form might give insight into the processes involved.
\end{itemize}
A perfect fluid has a gauge description \cite{mdr27};
this allows it to be canonically quantized producing a quantum fluid theory
which has novel quantum algebra,  see \cite{mdr27} and \S\ref{fbrane}.

To describe the approach here from the technical point of view:
if one considers a perfect fluid with stress
\be
T_{\mu\nu}=(\pr+\dn)V_\mu V_\nu+\pr g_{\mu\nu}.
\lb{1.6}
\ee
the $V_\mu V_\nu$ is a geometric object called the first fundamental
form of a one-dimensional surface,
which can also be though of as the tangent vector to the path of a point particle;
in general the first fundamental form of a $p+1$
surface and metric and projection tensor are equated by
\be
h^{\mu\nu}\equiv g^{\mu\nu}-\ff^{\mu\nu},~~~~~
h=d-1-p.
\lb{2.3}
\ee
$\ff^{\mu\nu}$ being the generalization of $V^\nu V^\mu$
where now the tangents are to a membrane,
so the question arises as to what a fluid with $p+1$ dimensional $\ff$ replacing the $V$'s
would look like and this is approached here through the lagrangian method.

In \S\ref{slpf} the properties of a perfect fluid are described.
In \S\ref{mem} some of the properties of Dirac membranes are described,
the treatment of the constraints here is a different from usual.
In \S\ref{ffluid} the lagrangian description of a perfect fluid is much generalized,
this involves putting internal indices on fluid objects,
the indices label the object and do not usually involve differentiation.
Up to adding similarly structured terms,
it is hoped that the resulting $f$-fluid is the most general
that can be derived by a spin-free lagrange method;
however it is not general enough to incorporate heat conduction or anisotropic stresses.
In \S\ref{fbrane} it is shown how to reduce the number of indices by using an internal metric,
this produces the $f$-brane for which both lagrangian and metric stress have both the perfect
fluid and membranes as examples.
Reduction is taken to have happened when both the lagrangian and metric stress coincide.

The notation used is:
signature (-+++),
greek indices $\al,\bt,\dots,\mu,\dots=0,1,2,3$ are spacetime indices,
early latin indices $a,b,c,\dots$ are internal or fluid indices,
middle greek indices $\iota\dots$ index constraints if they are not otherwise indexed,
middle latin indices $i,j,k,\dots$ are velocity potential indices,
all indices are left out when it is hoped that the ellipsis is clear,
$\ga$ is the auxiliary metric which is usually that of the internal space of a membrane,
$\pr$ is the pressure but $p+1$ the dimensions of $\ga$,
$\dn$ is the density, $n$ is the particle number,
$\en$ is the enthalpy but $h$ is the trace of the projection tensor
and $\hb$ is Planck's constant divided by $2\pi$,
$W$ is a vector field which has been decomposed into clebsch velocity potentials,
$V$ is a unit timelike vector field constructed from $W$ and the enthalpy $\en$.
All other conventions are those of Hawking and Ellis \cite{HE}.
\section{The Perfect Fluid.}
\label{slpf}
In this section the lagrangian formulation of a perfect fluid \cite{mdr27} is recalled.
The lagrangian of a perfect fluid is taken to be the pressure
\be
\Lg=\pr,
\lb{1.1}
\ee
and the hamiltonian is the density $\dn$.
The lagrangian (\ref{1.1}) is varied via the first law of thermodynamics
\be
{\rm d}\pr=n{\rm d}\en-nT{\rm d}s,
\lb{1.2}
\ee
where $n$ is the particle number,  $\en$ is the enthalpy,  $T$ is the temperature,
and $s$ is the entropy.  The fluid velocity vector $V$ is a unit timelike vector field
which has a clebsch decomposition
\be
\en V_\mu=W_\mu=\si_\mu+\te s_\mu+\dots,~~~~~~
V^\mu V_\mu=-1,~~~~~~
h_{\mu\nu}=g_{\mu\nu}+V_\mu V_\nu,
\lb{1.3}
\ee
where $\si,\te,s$ are the clebsch potentials,  $\si$ is the higgs \cite{mdr25},
$\te$ is the thermasy,  and $s$ the entropy.
The number of extra terms depends on the dimension of the spacetime involved.
clebsch's theorem is only local,  so that when there are obstructions to the
velocity $V$ or unusual global properties it no longer holds;
the usual way around this is to add yet more terms to the decomposition.
For our purposes when the first three terms are known it is straightforward
to add the extra terms,  so that we use just three terms regardless of
dimension and so on.  Using the unit normalization of $V$,  $V^2=-1$ and $\en V=W$
the first law (\ref{1.2}) can be written
\be
{\rm d}\pr=-nV_\mu{\rm d}W^\mu-nT{\rm d}s,
\lb{1.4}
\ee
using
\be
\pr+\dn=n\en,
\lb{1.5}
\ee
and varying with respect to the metric gives the metric stress \ref{1.6}.
The reduction to a congruence of point particles is achieved using
\be
\Lg=\pr=-m\el=-m\sq{-\dot{x}^2},~~
\en=\pm\el,n=\pm m,~~
\Hm=\dn=0,~~
V^\mu=\fr{\dot{x}^\mu}{\el},~~
T_{\mu\nu}=m\el h_{\mu\nu}.
\lb{1.7}
\ee
The point particles equation of motion is $\dot{P}/m=\dot{V}=0$ is not explicitly recovered,
$\dot{V}^a=V_bV^a_{~;b}$ is just the acceleration of the fluid,
so that in addition to (\ref{1.7}) the reduction requires that the fluid is acceleration free.
The reduction to a single point particle is achieved by either using delta functions
or by varying the lagrangian with respect to the single internal index,  see \cite{mdr33}.
Variation of (\ref{1.1}) with respect to the clebsch potentials give their equations of motion
\be
(nV^\mu)_{;\mu}=\dot{n}+n\Te=0,~~~
\dot{s}=0,~~~
\dot{\te}=T.
\lb{1.8}
\ee
The momenta are
\be
\Pi^\si=-n,~~~
\Pi^\te=0,~~~
\Pi^s=-n\te,
\lb{1.9}
\ee
and the constraints between the momenta are
\be
\ph_1=\Pi^s-\te\Pi^\si,~~~
\ph_2=\Pi^\te.
\lb{1.10}
\ee
The Poisson bracket is defined by
\be
\{A,B\}\equiv\fr{\de A}{\de q_i}\fr{\de B}{\de \Pi^i}-\fr{\de B}{\de q_i}\fr{\de A}{\de \Pi^i}.
\lb{1.11}
\ee
The Dirac matrix is defined by
\be
C_{\iota\kappa}\equiv\{\ph_\iota,\ph_\kappa\},
\lb{1.12}
\ee
and the Dirac bracket is defined by
\be
\{A,B\}*\equiv\{A,B\}-\{A,\ph_\iota\}C^{-1}_{\iota\kappa}\{\ph_\kappa,B\}.
\lb{1.13}
\ee
Quantization is achieved using the substitution of the Dirac bracket by commutators
\be
\{A,B\}*\rightarrow\fr{1}{i\hb}[\hat{A}\hat{B}-\hat{B}\hat{A}].
\lb{1.14}
\ee
For the perfect fluid
\be
C_{12}=\{\ph_1,\ph_2\}=\Pi^\si,~~~
C_{\iota\kappa}=-i\si^2\Pi^\si,~~~
C^{-1}_{\iota\kappa}=\fr{i\si^2}{\Pi^\si},~~~
\lb{1.15}
\ee
where $\si^2$ is the Pauli matrix
\ber
\si^2=
\left(\begin{array}{cc}
    0  & -i  \\
    i  &  0
\end{array}   \right).
\lb{1.16}
\ear
Forming the Dirac brackets between the fields and momenta
it turns out that it is possible to multiply by $\Pi^\si$ throughout.
After applying the quantization (\ref{1.14}) and dropping the hats
the equations between the fields and momenta are
\ber
&[\si\te-\te\si]\Pi^\si=\Pi^\si[\si\te-\te\si]=-i\hb\te\de^4(x-y),
[\te s-s\te]\Pi^\si=\Pi^\si[\te s-s\te]=-i\hb\de^4(x-y),\n
&[\Pi^\si\si-\si\Pi^\si]=-i\hb\de^4(x-y),
[\Pi^s-s\Pi^s]=-i\hb\de^4(x-y).
\lb{1.17}
\ear
Using the substitutions
\be
v_1=\si,~
v_2=s,~
v_3=\te,~
v_4=\Pi^\si,~
v_5=\Pi^s,~
v_6=\Pi^\te,
\lb{1.18}
\ee
setting $\hb=1$,  and then suppressing the delta function $\de^4(x-y)$ gives
\be
v_4(v_3v_2-v_2v_3)=-i,
v_4(v_1v_3-v_3v_1)=-iv_3,
v_4v_1-v_1v_4=-i,
v_5v_2-v_2v_5=-i.
\lb{1.19}
\ee
$v_6$ does not occur.
In the case when $V$ is a gradient vector $V^\mu=\si^\mu$,
only the second to last of these commutators remains,
so that the algebra is the same as that of the point particle.
\section{The Dirac Membrane.}\lb{mem}
The membrane lagrangian is \cite{dirac}
\be
\Lg=k\sq{-\ga},~~~
\ga_{ab}=x_{\mu a}x^\mu_b,~~~
\sq{-\ga}=(-\det\ga_{ab})^\fr{1}{2}.
\lb{2.1}
\ee
In order to discuss variation of (\ref{2.1}) it is necessary to introduce
the geometric objects the projection tensor (\ref{2.3}) and the first fundamental form
\be
\ff^{\mu\nu}=\ga^{ab}x^\mu_ax^\nu_b,~~~
\ff^{\mu\rh}\ff^\nu_\rh=\ff^{\mu\nu},~~~
\ff^\rh_\rh=\ga^c_c=x^{\rh c}x_{\rh c}=p+1.
\lb{2.2}
\ee
Varying the lagrangian (\ref{2.1}) with respect to the metric gives the metric stress
\be
T^{\mu\nu}=k\sq{-\ga}(-(p+1)\ff^{\mu\nu}+g^{\mu\nu}).
\lb{2.4}
\ee
Defining momenta by varying the lagrangian (\ref{2.1}) with respect to velocities
gives momenta and the constraints between
\be
P^a_\mu=-k\sq{-\ga}x^a_\mu,~~~
\ph^{ab}_{\mu\nu}\equiv P^a_\mu P^b_\nu+k^2\ga x^a_\mu x^b_\nu,
\lb{2.6}
\ee
compare \cite{pavsic} and \cite{cartas},
the constraints in (\ref{2.6}) have contractions
\ber
&\ph_{\mu\nu}=\ph^a_{a\mu\nu}=P^a_\mu P_{\nu a}+k^2\ga\ff_{\mu\nu},~~~
\ph^{ab}=\ph^{ab\mu}_{~~~\mu}=P^{a\mu}P^b_\mu+k^2\ga\ga^{ab},\n
&\ph=\ph^\mu_\mu=\ph^a_a=P^{a\mu}P_{a\mu}+k^2(p+1)\ga.
\lb{2.7}
\ear
Using just the last of these it is possible to derive a Klein-Gordon equation
with $k^2=m^2$,  see \cite{mdr42};  however in general one needs the other constraints.
Defining
\ber
&X^{ab\mu}_c\equiv\fr{\de \ph^{ab}}{\de P^{c\mu}}=2\de^{(a}_cP^{b)\mu}\\
&Y^{ab}_\mu\equiv\fr{\de \ph^{ab}}{\de x^\mu}
=-2k^2(\ga x^{(a}_\mu)^b-2k^2(\ga\ga^{ab}x^c_\mu)_c
=-2k\left(\sq{-\ga}(P^c_\mu\ga^{ab}+P^{(a}_\mu\ga^{b)c}\right)_c.\nonumber
\lb{2.8}
\ear
$X$ has one more index than $Y$,
so that it is not possible to form a Dirac matrix $C$ for them as they are.
Choosing $c=\ta$ in $X$ one can form $C_{abcd}=\{\ph_{ab},\ph_{cd}\}$,
but it is not clear what the inverse of this should be.
Choosing $b=c=\ta$ in $X$ and $b=\ta$ in $Y$
one sees that the constraints are limited to the
$(p+1)$ first class constraints $\ph^a=\ph^{a\ta}$
and one can form Dirac matrix $C_{ab}=\{\ph_a,\ph_b\}$.
Restricting to the case of the string $p=1$ and
\be
C_{ab}=i\si^2C_{\ta\si}=i\si^2Z^{\ta\si\mu}_{~~~~\mu},~~~
C^{-1}_{ab}=\fr{-i\si^2}{Z^{\ta\si\mu}_{~~~~\mu}},~~~
Z^{\ta\si}_{\mu\nu}
\equiv Y^{\ta\ta}_\nu X^{\si\ta}_{\ta\mu}-Y^{\si\ta}_\nu X^{\ta\ta}_{\ta\mu}.
\lb{2.9}
\ee
The Dirac brackets between the coordinates and momenta are
\be
\{x^\mu,x^\nu\}*=\{P^\mu,P^\nu\}*=0,~~~
\{x^\mu,P^\nu\}*=g^{\mu\nu}-i\si^2\fr{Z^{\ta\si\mu\nu}}{Z^{\ta\si\rh}_{~~~~\rh}}.
\lb{2.10}
\ee
One can get an explicit form of these Dirac brackets by using a specific form of the internal
metric $\ga$,  say the Nambu-Goto choice,  see for example \cite{mdr33},
however from the present perspective the important point is that now the $\{x,P\}*$ Dirac
bracket is different from the Poisson bracket,  whereas for the perfect fluid the
$\{q,q\}*$ bracket also differs.
\section{The F-fluid.}\lb{ffluid}
The lagrangian (\ref{1.1}) is generalized to depend on $F$ presuures
\be
\Lg=f(\pr_1,\pr_2,\dots,\pr_F),
\lb{3.1}
\ee
the pressures $\pr$ and densities $\dn$ are
equated to the enthalpies $\en$ and particle numbers $n$ by
\be
\pr_a+\dn_a=n^{bc}_a\en_{bc},
\lb{3.2}
\ee
which generalizes (\ref{1.3}) the internal indices $a,b,\dots$ label distinct objects,
for example $\pr_1,\pr_2\dots$ and only indicate differentiation
with respect to the index when it is on $x$ or is outside a bracket $()_a$.
The unit timelike condition on $V$ (\ref{1.3}) generalizes to
\be
h^c_b V^b_\mu=W^c_\mu,~~~
V^\mu_d V^e_\mu=-\de^e_d,
\lb{3.3}
\ee
so that
\be
-\en_{ab}=-\de^c_b \en_{ac}=\en_{ac} V^\mu_b V^c_\mu=V^\mu_b W_{a\mu}.
\lb{3.4}
\ee
(\ref{3.4}) allows the first law (\ref{1.2}) to be generalized to
\be
{\rm d}p_a
=n^{bc}_a{\rm d}\en_{bc}-n^{bc}_aT_b{\rm d}s_c
=-n^{bc}_aV_{\mu b}{\rm d}W^\mu_c-n^{bc}_aT_b{\rm d}s_c.
\lb{3.5}
\ee
Varying (\ref{3.1}) with respect to the spacetime metric gives the metric stress
\be
T_{\mu\nu}=f_{,\pr_a}n^{bc}_a \en_{ce}V_{\mu b}V^e_\nu+fg_{\mu\nu}.
\lb{3.6}
\ee
The clebsch decomposition (\ref{1.3}) of $W$ generalizes to
\be
W^c_\mu=\si^c_\mu+\al^c_{ab}\te^as^b_\mu+\dots.
\lb{3.7}
\ee
In the present case this decomposition is an assumption,
unlike for the perfect fluid where it is a consequence of Clebsch's theorem,
$\al^c_{ab}$ is a new object.
Varying with respect to the potential gives the equations of motion
\be
(V^\mu_bn^{bc}_a)_\mu=0,~~~
n^{bc}_a\al_{cde}V^\mu_bs^e_\mu=0,~~~
V^\mu_bn^{bc}_a(\al_{cde}\te^d)_\mu-n^{be}_aT_b=0,
\lb{3.8}
\ee
for $\si^a,~\te^a,~s^a$ respectively,  compare (\ref{1.8}).
The momenta generalizing (\ref{1.9}) are
\be
\Pi^{\si bc}_{~~~~a}=-n^{bc}_a,~~~
\Pi^\te_a=0,~~~
\pi^{sbc}_{~~~~a}=-n^{bc}_a\al\cdot\te,
\lb{3.9}
\ee
where $\al_a=\al^c_{ac}$.
The constraints generalizing (\ref{1.10}) are
\be
\ph^{bc}_a\equiv\Pi^{sbc}_a-\al\cdot\te\Pi^{\si bc}_{~~~~a},~~~
\ph^2_a\equiv\Pi^\te_a.
\lb{3.10}
\ee
In this form it is not possible to get an inverse Dirac matrix.
\section{The F-brane.}\lb{fbrane}
The $f$-brane is a particular case of the $f$-fluid.
First reduce the number of indices by
\be
n^{bc}_a=\ga^{bc}n_a,~~~
\en_{bc}=\fr{1}{p+1}\en\ga_{bc},
\lb{4.1}
\ee
where $\ga$ is an internal metric,  this gives
\be
\pr_a+\dn_a=\en n_a,~~~
\en V^c_\mu=W^c_\mu,~~~
V^c_\mu V^\mu_e=-\de^c_e.
\lb{4.2}
\ee
The first law is
\be
{\rm d}p_a
=n_a\ga^{bc}{\rm d}(\en\ga_{bc})-\ga^{nc}n_aT_b{\rm d}s_c
=-n_aV^c_\mu{\rm d}W^\mu_c-n_aT^c{\rm d}s_c,
\lb{4.3}
\ee
where one can take $\ga$ through the d to get the second equality.
The metric stress is
\be
T_{\mu\nu}=f_{,\pr_a}n_a\en V^c_\mu V_{c\mu}+fg_{\mu\nu}.
\lb{4.4}
\ee
The best identification of $V$ is not immediate.
The problem is that for the point particle the explicit normalization condition is
$V_{\mu\ta}V^\mu_\ta=-1$,
with $\ta$ subscripted both times,
for only one internal index this does not matter,
but for more than one internal index one wants to preserve covariance,
so that a different choice has to be made,  choosing
\be
V^{\mu c}=\bt x^{\mu c},
\lb{4.5}
\ee
the normalization condition (\ref{4.1}) and the definition of $\ga$ require
\be
V^c_\mu V^\mu_e=\bt^2x^{\mu c}x_{\mu e}=\bt^2\ga^c_e=-\de^c_e
\Rightarrow\bt=i.
\lb{4.6}
\ee
With this choice it is possible to introduce the first fundamental form into the metric stress
\be
T_{\mu\nu}=-f_{,\pr_a}n_a\en\ff_{\mu\nu}+fg_{\mu\nu},
\lb{4.7}
\ee
choosing the simplest lagrangian
\be
\Lg=f=\Si \pr_a,
\lb{4.8}
\ee
gives the metric stress of the $f$-brane
\be
T_{\mu\nu}=-\Si(\pr_a+\dn_a)\ff_{\mu\nu}+\Si\pr_ag_{\mu\nu}.
\lb{4.9}
\ee
This is much as expected,
the only other possibilities for its form would be terms summing the internal indices
occurring across the pressure part and the $x$ part.
The $f$-brane reduces to the Dirac membrane when
\be
\Si\pr_a=k\sq{-\ga},~~~
\Si\dn_a=pk\sq{-\ga},
\lb{4.10}
\ee
different choices can reduce to the stress of the conformal membrane \cite{mdr33}\S4.
Varying with respect to the potentials the equations of motion are
\be
(V^{\mu c}n_a)=0,~~~
n_a\al^b_{de}V^\mu_bs^e_\mu=0,~~~
V^{\mu c}(\al_{cde}\te^d)_\mu-T^e=0.
\lb{4.11}
\ee
The momenta are as for the $f$-fluid,
however choosing $\Pi^\si_a\ga^{bc}=\Pi^{\si bc}_{a}$ and similarly for $\Pi^s$ gives
\be
\Pi^\si_a=-n_a,~~~
\Pi^\te_a=0,~~~
\Pi^s_a=-\al\cdot\te n_a,
\lb{4.12}
\ee
which have constraints
\be
\ph^1_a\equiv\Pi^s_a-\al\cdot\te\Pi^\si_a,~~~
\ph^2_a\equiv\Pi^\te_a.
\lb{4.13}
\ee
These constraints give Dirac bracket
\be
C^{12}_{ab}=\{\ph^1_a,\ph^2_b\}=\al_b\Pi^\si_a.
\lb{4.14}
\ee
To proceed it is necessary to find an inverse of this.
The simplest way is to reduce it from a four indexed matrix to a two index matrix by
tracing across the internal indices
\be
C^{a~12}_{~a}=\{\ph^{1a},\ph^2_a\}=\al\cdot\Pi^\si,
\lb{4.15}
\ee
and then proceed as for the perfect fluid in \S\ref{slpf}.
Defining $v$'s as in (\ref{1.18}),
except that this time they all are subscripted by an internal index,
the following algebra is found
\ber
\label{4.16}
&\al\cdot v^4(v^1_av^3_b-v^3_bv^1_a)=-i\al\cdot v^3\de_{ab},~~~
\al\cdot v^4(v^3_av^2_b-v^2_bv^3_a)=-i\de_{ab},\\
&\al\cdot v^4(v^3_av^6_b-v^6_bv^3_a)=-i(\al\cdot v^4\de_{ab}-\al_b v^4_a),~~~
v^4_av^1_b-v^1_bv^4_a=-i\de_{ab},~~~
v^5_av^2_b-v^2_bv^5_a=-i\de_{ab}.\nonumber
\ear
which reduces to (\ref{1.19}) when $a,b,\dots=1$.
Note that $v^6$ occurs in (\ref{4.16}) but not in (\ref{1.19}).
\section{Conclusion.}\lb{conc}
Technically the
classical perfect fluid can be reduced to a congruence of point particles using
the reduction equations (\ref{1.7}).
The perfect fluid has one propagating degree of freedom,  the particle number,
and two independent functions in the stress,  the pressure and density;
whereas the congruence of point particles has one propagating degree of freedom,
and one independent function in the stress.
The corresponding quantum theory reduces when the velocity vector $W$ is a gradient vector.
In the classical case a perfect fluid can be generalized to
a $f$-brane (\ref{4.8}),(\ref{4.9}) which reduces to the Dirac membrane.
The $f$-brane has $p+1$ propagating degrees of freedom,  the $p+1$ particle numbers,
and $2p+2$ independent functions in the stress.
There is no indication that this correspondence remains so quantum mechanically,
this is because:
the fluid momenta lack a spacetime index,
whereas membrane momenta have a spacetime index;
also for membranes it is the $\{\Pi,x\}*$ Dirac bracket which differs from the Poisson bracket,
whereas for fluids the $\{q,q\}$ brackets differs as well.

The prospects for contact with other theories can be guessed.
The possibility of contact with M-theory depends on finding some way of the calculating quantum
states of the theory,  but how this could be achieved is unknown.
In principle one could replace all the partial derivatives $\p_\mu\rightarrow\p_\mu+ieA_\mu$
to 'charge' the theory and discuss symmetry breaking but it is not clear yet where this
could lead.
Contact with existing fluid models of low temperature physics would require addition of more
thermodynamical objects and it is not clear how this could be achieved in a lagrangian approach
to a perfect fluid,  there is the possibility that the more general fluids presented here might
have emergent thermodynamical properties in some new reduction.
The fluids presented here could be coupled to field equations such as those of general relativity
in the hope of producing new quantum cosmologies,
but it is hard to envisage what this would achieve.
In many particle processes one could take the internal indices to correspond
to a species of particle,  this could lead to the embedded space being
of higher dimension than spacetime so that the geometric interpretation is lost.


\begin{thebibliography}{99}

\bibitem{AS}
A. Aurilia \& Euro Spallucci,
Gauge Theory of Relativistic Membranes.
Math.Rev.\href{http://www.ams.org/mathscinet-getitem?mr=94d:81139}
{\sl 94d:81139},
\href{http://arXiv.org/abs/hep-th/9305020}
                      {\tt hep-th/9305020}
Class.Quantum Grav.{\bf 10}(1993)1217-1248.

\bibitem{BD}
N.D. Birrel \& P.C.W. Davies,P.C.W.
Quantum Fields in Curved Space.
Cambridge University Press (1982),
The Pitt Building,  Trumpington Street,  Cambridge CB2 1RP,  UK.

\bibitem{cartas}
R. Cartas-Fuentevilla,
Indentically closed two-form for covariant phase space quantization
of the Dirac-Nambu-Goto p-brane in a curved spacetime.
\href{http://arXiv.org/abs/hep-th/0204133}
                      {\tt hep-th/0204133}
{\it Phys.Lett.}{\bf B536}(2002)283-288.

\bibitem{CK}
B. Carter \& I.M. Khalatnikov,
Momentum, vorticity, and helicity in covariant superfluid dynamics.
Math.Rev.\href{http://www.ams.org/mathscinet-getitem?mr=93m:83046}
{\sl 93m:83046}
{\it Ann.Physics}{\bf 219}(1992)243-265.

\bibitem{dirac}
P.A.M. Dirac,
An Extendible Model of the Electron.
{\it Proc.Roy.Soc.}{\bf A268}(1962)57-67.

\bibitem{GHY}
G. Gibbons, Kentaro Hori \&  Piljin Yi,
String Fluid from Unstable D-branes.
{\it Nucl.Phys.}{\bf B596}(2001)136-150.
\href{http://arXiv.org/abs/hep-th/0009061}
                      {\tt hep-th/0009061}

\bibitem{GSW}
M.B. Green,  J.H.  Schwarz \& E. Witten,
Superstring Theory:  Volume 1 Introduction.
Cambridge University Press (1987).

\bibitem{GS}
E.I. Guendelman \& E. Spallucci,
Conformally invariant gauge theory of three-branes in 6-D
and the cosmological constant,
{\it Phys.Rev.D}{\bf 70}(2004)026003.
\href{http://arxiv.org/abs/hep-th/0311102}
                      {\tt hep-th/0311102}

\bibitem{HE}
S.W. Hawking \& G.F.R. Ellis (1973)
{\it The Large Scale Structure of Space-time.}
Math.Rev.\href{http://www.ams.org/mathscinet-getitem?mr=54:12154}
{\sl 54 \#12154},  Cambridge University Press,
The Pitt Building,  Trumpington Street,  Cambridge CB2 1RP,  UK.

\bibitem{HL85}
W.A. Hiscock \& L. Lindblom,
{\it Phys.Rev.}{\bf D 31}(1985)725.

\bibitem{hoppe}
Jens Hoppe,
Surface Motions and Fluid Dynamics.
\href{http://arXiv.org/abs/hep-th/9405001}
                      {\tt hep-th/9405001}
{\it Phys.Lett.}{\bf B335}(1994)41-44.

\bibitem{j1}
R. Jackiw,
A Particle Field Theorist's Lectures on Supersymmetric,
Non-Abelian Fluid Mechanics and d-Branes.
\href{http://arXiv.org/abs/physics/0010042}
                      {\tt hep-th/0010042}

\bibitem{bi:LR}
V.G. Lapchinskii \&  V.A. Rubakov,
{\it Theor.Math.Phys.}{\bf 33}(1977)1076.

\bibitem{pavsic}
Matej Pav\v si\v c,
Phase Space Action for Minimal Surfaces of any Dimension in Curved Spacetime.
Math.Rev.\href{http://www.ams.org/mathscinet-getitem?mr=89b:83016}
{\sl 89b:83016}
{\it Phys.Lett.}{\bf B197}(1987)327-331.

\bibitem{mdr25}
Mark D. Roberts,
Fluid Symmetry Breaking II: Velocity Potential Method.
{\it Hadronic J.}{\bf 20}(1997)73-84.
\href{http://arXiv.org/abs/hep-th/9904079}
                      {\tt hep-th/9904079}

\bibitem{mdr27}
Mark D. Roberts,
The Quantum Commutator Algebra of a Perfect Fluid.
{\it Mathematical Physics,  Analysis and Geometry}{\bf 1}(1999)367-373.
\href{http://arXiv.org/abs/gr-qc/9810089}
                      {\tt gr-qc/9810089}

\bibitem{mdr33}
Mark D. Roberts,
The Rotation and Shear of a String.
{\it Class.Q.Grav.}{\bf 20}(2003)507-519.
\href{http://arXiv.org/abs/hep-th/0204236}
                      {\tt hep-th/0204236}

\bibitem{mdr42}
Mark D. Roberts,
The Relative Motion of Membranes.
{\it Central European Journal of Physics}{\bf 8}(2010)915-919.
\href{http://arXiv.org/abs/gr-qc/0404094}
                      {\tt gr-qc/0404094}

\bibitem{mdr0910}
Mark D. Roberts,
The clebsch potential approach to fluid lagrangians.
\href{http://arXiv.org/abs/0910.3587}
                      {\tt 0910.3587}

\bibitem{tseytlin}
A.A. Tseytlin,
Sigma Model Approach to String Theory.
Math.Rev.\href{http://www.ams.org/mathscinet-getitem?mr=91c:81117}
{\sl 91c:81117}
{\it Int.J.Mod.Phys.}{\bf A4}(1989)1257.

\end{thebibliography}
\end{document}